\documentclass[usenatbib,referee]{ctextemp_mn2e}

\title[The progenitor of VFTS102]
{The binary merger channel for the progenitor of the fastest
rotating O-type star VFTS102}

\author[Jiang et al.]{Dengkai Jiang$^{1,2}$\thanks{E-mail:
dengkai@ynao.ac.cn}, Zhanwen Han$^{1,2}$, Liheng Yang$^{3}$ and Lifang Li$^{1,2}$\\
$^{1}$National Astronomical Observatories, Yunnan Observatory,
Chinese Academy of Sciences, P.O. Box 110,
Kunming,\\
\ \ \ \  \ \ \ \ \ \ \ \ \ \ \ \ \ \ \ \ \ \ \ \ \ \ \ \ \ \ \ \ \
\ \ \ \  Yunnan Province, 650011, P.R. China\\
$^{2}$Key Laboratory for the Structure and Evolution of Celestial
Objects, Chinese Academy of Sciences
\\
$^{3}$National Astronomical Observatories,Chinese Academy of
Sciences, Beijing 100012, China}
\begin{document}
\input ctextemp_psfig.sty
\date{Accepted .... Received .....; in original form ....}

\pagerange{\pageref{firstpage}--\pageref{lastpage}} \pubyear{2009}

\maketitle

\label{firstpage}

\begin{abstract}
VFTS102 has a projected rotational velocity ($>$500\,km s$^{-1}$)
and would appear to be the fastest rotating O-type star. We show
that its high rotational velocity could be understood within the
framework of the binary merger. In the binary merger channel, the
progenitor binary of VFTS102 would evolve into contact while two
components are still on the main sequence, and then merge into a
rapidly rotating single star. Employing Eggleton's stellar evolution
code, we performed binary stellar evolution calculations and mapped
out the initial parameters of the progenitor of VFTS102 in the
orbital period-mass ratio ($P-q$) plane. We found that the
progenitor binary of VFTS102 with initial mass ratio $q_0 \la 0.7$
should have an initial orbital period shorter than
$3.76-4.25$\,days, while above this mass ratio it should have an
initial orbital period shorter than $1.44-1.55$\,days. The
progenitor of VFTS102 would evolve into contact during the rapid
mass transfer phase or during the subsequent slow mass transfer
phase, and might ultimately merge into a rapidly rotating massive
star. In addition, we performed Monte Carlo simulations to
investigate the binary merger channel. We estimated the fraction of
binaries that would merge into single stars and the fraction of
single stars that might be produced from the binary merger channel.
It is found that about 8.7\% of binaries would evolve into contact
and merge into rapidly rotating single stars, and about 17.1\% of
single stars might be produced from the binary merger channel and
should have similar properties to VFTS102. This suggests that the
binary merger channel might be one of the main channels for the
formation of rapidly rotating massive stars like VFTS102.

\end{abstract}

\begin{keywords}
instabilities -- stars: early-type -- stars: formation -- stars:
evolution -- stars: rotation
\end{keywords}

\section{Introduction}
VFTS102 is a rapidly rotating O-type star in the 30 Doradus region
of the Large Magellanic Cloud, and its projected rotational velocity
is larger than 500\,km s$^{-1}$ and probably as large as 600\,km
s$^{-1}$ \citep{Dufton 2011}. It rotates more rapidly than any
observed stars in recent large surveys \citep{Martayan 2006, Hunter
2008, Hunter 2009} and would appear to be the most rapidly rotating
massive star \citep{Dufton 2011}. The rapidly rotating stars such as
VFTS102 form an important class of objects in several respects.
These stars could help us to study the rotational effects that are
important, from star formation through to their deaths \citep{Hunter
2008}. Fast rotation could change the lifetime and evolution of
massive stars by rotationally induced mixing between the core and
envelope \citep{Heger 2000, Meynet 2000}, and produce enrichment of
a number of different elements at the stellar surface \citep{Hunter
2009, Frischknecht 2010, Potter 2012}. The rotationally induced
mixing were used to explain the variety of core collapse supernovae
\citep{Georgy 2009}. In addition, the rotating massive stars might
be the progenitors of gamma-ray bursts through homogeneous evolution
\citep{Yoon 2005, Woosley 2006, Tout 2011}.

Two channels for the formation of VFTS102 have been proposed in the
past. The first channel is the mass transfer channel \citep{Dufton
2011}. In this channel, VFTS102 might be the mass gainer and be spun
up by a past episode of Roche lobe overflow (RLOF) in an interacting
binary system \citep{Packet 1981, Cantiello 2007, Dufton 2011}.
\citet{Dufton 2011} suggested that VFTS102 became a runaway star
after its companion exploded as a supernova. This channel could also
explain that the radial velocity of VFTS102 differs by 40\,km
s$^{-1}$ from the mean radial velocity for 30 Doradus. The second
channel is the collision channel \citep{Fryer 2005, Fujii 2012}.
\citet{Fujii 2012} proposed that VFTS102 had experienced a earlier
collision, and then might be ejected from the cluster center. The
fast rotation of VFTS102 might be the result of the collision with
another star in the cluster center. In addition, VFTS102 might be
born as rapid rotator \citep{Huang 2010, Brott 2011} or its surface
might spin up during the main sequence (MS) evolution because
angular momentum was transported from the center to the stellar
surface \citep{Ekstrom 2008}.

From a binary evolution point of view, these formation channels are
not complete. The binary merger might be another possible channel to
produce the fastest rotating star VFTS102. In this channel, the
initial detached binary would experience case A mass transfer and
evolve into contact while two components are still on the MS, and
then merge into a rapidly rotating single star \citep{Webbink 1976,
Jiang 2010, de Mink 2011, Langer 2012}. It is well known that the
fraction of massive stars that are members of a close binary is very
large \citep{Mason 2009, Langer 2012}. \citet{Sana 2012} found that
20\% of all stars born as O-type stars would merge as a result of
case A mass transfer. In addition, the observations show that the
massive contact binaries are relatively common, such as V606 Cen
\citep[$P$=1.495\,d][]{Lorenz 1999} and TU Mus
\citep[$P$=1.387\,d][]{Terrell 2003, Qian 2007}, and $\sim 5\%$ of
O-type stars appear to be in contact with orbital periods of $\sim
1-5.5$\,days \citep{Garmany 1980, Hilditch 1987, Eggleton 1996}.
These massive contact binaries might merge into single stars in a
similar way as low-mass contact binaries, which might merge into
single stars due to tidal instability when the spin angular momentum
of the system is more than a third of its orbital angular momentum
\citep{Hut 1980, Rasio 1995, Li 2006}, or due to thermal instability
when the primary attempts to cross the Hertzsprung gap
\citep{Webbink 1976}. The merger of contact binaries are expected to
result in single, massive stars \citep{Chen 2008, Tout 2011}, which
would be extremely rapidly rotating due to the orbital angular
momentum of the binary systems \citep{Cantiello 2007, Jiang 2010, de
Mink 2011, Langer 2012}. Therefore, the binary merger might be a
possible channel for the production of VFTS102.

The evolution of massive close binaries has been investigated by
many authors, e.g. \citet{Podsiadlowski 1992, Pols 1994, Wellstein
2001, Nelson 2001, de Mink 2007}. \citet{Nelson 2001} calculated the
case A binary evolution and constructed a large grid of models
(0.8$\leq M_{10} \leq 50$). They mapped out the initial parameters
in the mass ratio-orbital period plane of six subtypes of case A
evolution. They suggested that three of these subtypes (AR, AS, AD)
lead to contact while both components are on the MS. In case AR or
AS, the secondary expands in response to the thermal time-scale mass
transfer or the nuclear time-scale mass transfer from the primary
and fills its own Roche lobe. The system probably form a stable
contact binary, and might ultimately merge into a single star.
However, in case AD, the secondary cannot accrete all the proffered
material transfer from the primary on the dynamical time-scale. This
probably leads very quickly to a common envelope, spiral-in
\citep{Paczynski 1976, Hjellming 1987, Ge 2010}, and coalescence on
a quite short timescale \citep{Nelson 2001, Jiang 2012}. Case AD
might be common in the low mass binaries \citep{Nelson 2001, Jiang
2012}, especially in those binaries with two M dwarf components
observed by \citet{Becker 2011} and \citet{ Nefs 2012}. Few massive
systems are classified as AD and most of massive contact systems are
classified as AR or AS \citep{Nelson 2001, de Mink 2007}. Therefore,
the investigation of the evolution of massive binaries in case AR
and AS is important to understand the formation of massive contact
binaries, and then the formation of the rapidly rotating stars such
as VFTS102.

The purpose of this study is to investigate the binary merger
channel for the progenitor of VFTS102 and to determine the detailed
parameter range in which this channel produces VFTS102. Employing
the Eggleton's stellar evolution code, we construct a grid of binary
models for metallicity $Z = 0.01$ in Section 2 and 3, and then
implement the results in a binary population synthesis study in
Section 4. In Section 5, we give the discussion and conclusions.

\section{binary evolution calculations}
In the binary merger channel for the progenitor of the fastest
rotating O-type star VFTS102, the primary of the initial detached
binary would fill its Roche lobe and transfer some of its mass to
the secondary. The secondary expands in response to the mass
transfer from the primary. This system would evolve into contact if
the secondary fills its own Roche lobe when two components are still
MS stars. These contact systems would ultimately merge into rotating
stars due to the orbital angular momentum of the binary systems
\citep{Jiang 2010, Langer 2012}. Furthermore, if both components are
MS stars in case A binaries, the merged stars would be MS stars and
evolve in a similar way to a normal star with that mass \citep{Chen
2009, Langer 2012}. Therefore, the merged stars with the same mass
as VFTS102 might be similar to VFTS102.

There are some constraints on the progenitor of VFTS102 in the
binary merger model. First, the detached binary, as the progenitor
of VFTS102, should evolve into contact while two component are still
on the MS, which is largely determined by the initial period and
mass ratio. Secondly, the total mass of the progenitor binary should
be larger than or equal to the mass of VFTS102, which depends on the
mass loss during the evolution and the merger process. The current
mass of VFTS102 is approximately 25\,M$_{\rm \odot}$ \citep{Dufton
2011}. In this study, we do not consider the mass loss during the
evolution and merger process and assume that the binary system as
the progenitor of VFTS102 has total mass $\sim 24.5-25.5$\,M$_{\rm
\odot}$.

The merger process is complicated and the merger physics is still
uncertain. Here, we adopt the following assumptions: (i) the binary
systems merge immediately once both MS components fill their Roche
lobes; (ii) the merged stars are homogeneously mixed; (iii) the
system mass is conserved. The merged timescale (i.e. the time from a
binary contact to merger) for massive contact binaries has remained
unclear and we only adopt a simple assumption that the merger is
instantaneous. As the components of contact binaries rotate rapidly
during the merger process, the merged stars would be efficiently
mixed. These merged stars might also undergo rotationally induced
mixing during the subsequent evolution \citep{Langer 2012}, which is
important in the evolution of massive stars \citep[e.g.][]{Heger
2000, Maeder 2000, Howarth 2001}. Therefore, it is reasonable to
assume that the merged stars are homogeneously mixed. We roughly
assume that the mass is conservative during the merger process. In
fact, the binary systems might lose high angular momentum material
during the merger process and the merged stars might rotate at
subcritical rotation velocities \citep{Langer 2012}. The second and
the third assumptions were often adopted in the study of the
formation of blue stragglers by the binary merger
\citep[e.g.][]{Andronov 2006, Chen 2008, Chen 2009}.

To determine whether the detached binary evolves into contact when
two component are still MS stars, it is necessary to perform
detailed binary evolution calculations. Here we use Eggleton's
stellar evolution code originally developed by \citet{Eggleton 1971,
Eggleton 1972, Eggleton 1973}. This code has been updated with the
latest input physics during the last four decades \citep[e.g.][]{Han
1994, Pols 1995, Pols 1998, Nelson 2001, Eggleton 2002}. We
calculated our models with metallicity $Z = 0.01$ (hydrogen
abundance $X = 0.73$ and Helium abundance $Y = 0.26$), which is
close to the chemical composition of Large Magellanic Cloud
($Z=0.008$). The mass transfer is determined to be dynamic when the
mass-transfer is greater than $10M/t_{\rm KH}$, where $t_{\rm KH}$
is the thermal or Kelvin-Helmholtz timescale \citep{Nelson 2001}. We
construct a grid of stellar evolutionary models that covers the
following ranges of initial primary mass $M_{10}$ and initial mass
ratio ($q_0=M_{20}/M_{10}$):
\begin{equation}
\log M_{10} = 1.10, 1.15,..., 1.35,
\end{equation}
\begin{equation}
\log (1/q_{0}) = 0.004, 0.05, 0.10,..., 0.95,
\end{equation}
\begin{equation}
\log (P_0/P_{\rm ZAMS}) = 0.05, 0.10,..., 0.75,
\end{equation}
where $P_{\rm ZAMS}$ is the period at which the initially more
massive component would just fill its Roche lobe on the zero-age MS
\citep{Nelson 2001}. The orbits are assumed to be circular. A binary
is expected to become circularized during the RLOF by the tidal
force on a timescale which is much smaller than the nuclear
timescale \citep{Wang 2010}. We assume that both total mass and
orbital angular momentum are conserved in the evolution of the
binaries. If the loss of the total mass and orbital angular momentum
is considered, the orbit of the initial detached binaries would
quickly shrink. More detached binaries with longer period would
evolve into RLOF and contact, which could not evolve into contact in
the conservation evolution.

\begin{figure*}
\centerline{\psfig{figure=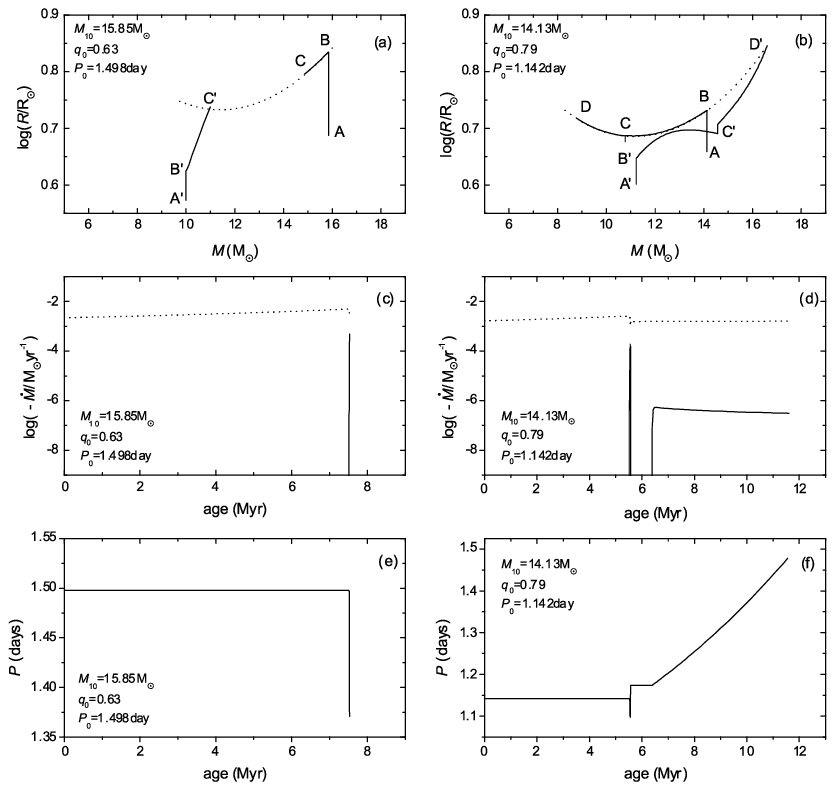,width=15.0cm}} \caption{Two
examples of binary evolution calculations. In panels (a) and (b),
solid curves show the evolutionary tracks in the radius-mass diagram
for the primaries (ABC and ABCD) and the secondaries (A'B'C' and
A'B'C'D'), and the dotted curves represent the Roche lobe. The
components start at point A or A', and evolve along curves until
both components fills their Roche lobe at point C and C', or D and
D'. In panels (c) and (d), the solid curves show the mass transfer
rates and the dotted curves represent 10$M/t_{\rm KH}$ that is used
to determined whether the mass transfer is dynamic. The evolution of
orbital period is shown as solid curves in panels (e) and (f).}
\label{fig1}
\end{figure*}

\begin{figure*}
\centerline{\psfig{figure=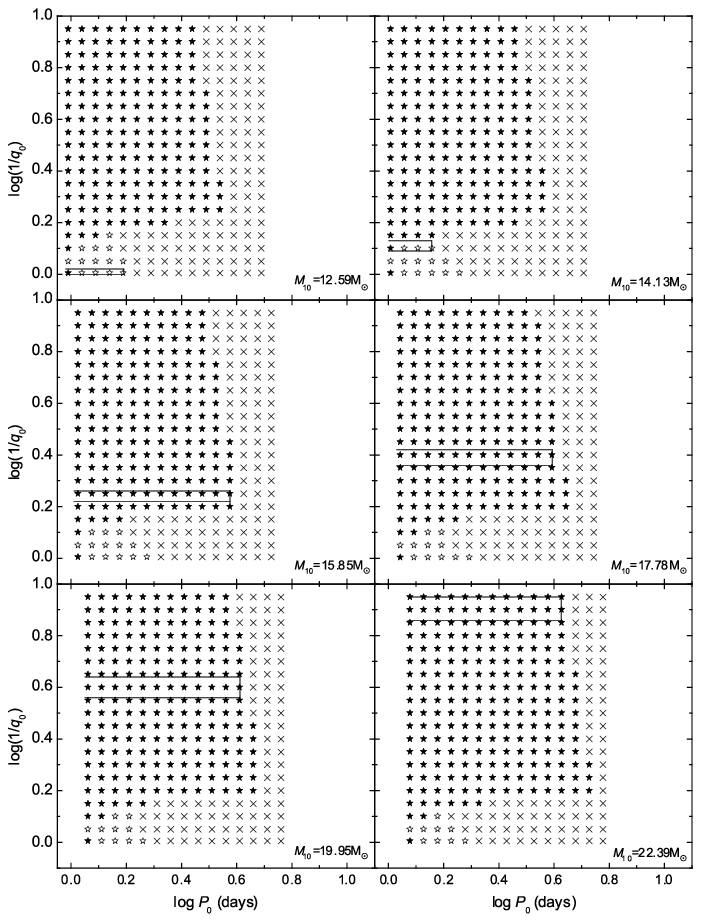,width=13.0cm}} \caption{The
outcomes of binary evolution calculations in the initial orbital
period$-$mass ratio ($\log P_0$, $\log (1/q_0)$) plane while both
components are still MS stars, where $P_{0}$ is the initial orbital
period and $q_{0}$ is the initial mass ratio (for different initial
primaries as indicated in each panel). Filled stars show the systems
that evolve into contact during the rapid mass transfer phase while
both components are on the MS (case AR). Open stars show the systems
that evolve into contact during the slow mass transfer phase (case
AS). Crosses indicate the systems that could not evolve into contact
while two components are still on MS. Solid lines show the
boundaries (without left boundaries) for the initial parameters for
which might be the progenitor of VFTS102.} \label{fig2}
\end{figure*}

\section{binary evolution results}

In Fig. 1, we present two representative examples of our binary
evolution calculations that could evolve into contact while two
components are still on the MS. It shows the evolutionary track in
the radius-mass diagram, the mass-transfer rate and the evolution of
the orbital period. Figs 1(a), (c) and (e) represent the evolution
of a binary system with an initial mass of the primary of $\log
M_{10}$ = 1.20 ($M_{10}$ =15.85\,M$_{\rm \odot}$), an initial mass
ratio of $\log (1/q_0)$=0.20 ($q_0 = 0.63$) and an initial orbital
period of $\log (P_0/P_{\rm ZAMS})= 0.2$ ($P_0 = 1.498$\,d). The
components start to evolve from point A and A', and the primary
fills its Roche lobe on the MS at point B which results in case A
RLOF (BC) as shown in Fig. 1(a). The secondary expands in response
to the mass transfer (B'C') and fills its Roche lobe at point C'.
This system evolves into contact shortly after the start of RLOF.
The mass-transfer rate from the primary to the secondary is not
greater than $10M/t_{\rm KH}$ as shown in Fig.1(c). Therefore, the
mass transfer is on a thermal time-scale and this case of binary
evolution is classed as case AR \citep{Nelson 2001}. Fig. 1(e) shows
that the orbital period of the system decreases from 1.498 to
1.37\,d during the mass transfer phase because the mass is
transferred from the massive component to the low-mass component.

Figs 1(b), (d) and (f) show another example for an initial mass of
the primary of $\log M_{10}$ = 1.15 ($M_{10}$ =14.13\,M$_{\rm
\odot}$), an initial mass ratio of $\log (1/q_0)$=0.10 ($q_0 =
0.79$) and an initial orbital period of $\log (P_0/P_{\rm ZAMS})=
0.1$ ($P_0 = 1.142$\,d). The main difference between this example
and the previous one is that the secondary fills its own Roche lobe,
not during the rapid mass transfer phase (B'C'), but during the
subsequent slow mass transfer phase (C'D') as shown in Fig. 1(b).
This case of binary evolution is identified as case AS \citep{Nelson
2001}. In this case, as shown in Fig. 1(d), the mass transfer rate
is about $1.8 \times 10^{-4}$\,M$_{\rm \odot}$/yr during the rapid
mass transfer phase (BC) and about $5 \times 10^{-7}$\,M$_{\rm
\odot}$/yr during the slow mass transfer phase (CD). The orbital
period of this system first decreases from 1.142 to 1.097d, and then
increases to 1.478\,d after the mass ratio is reversed as shown in
Fig. 1(f).

Fig. 2 summarizes the outcomes of binary evolution calculations in
the initial orbital period$-$mass ratio ($\log P_0$, $\log (1/q_0)$)
plane while both components are still MS stars. Filled stars show
the systems that evolve into contact during the rapid mass transfer
phase while both components are on the MS (case AR). Open stars show
the systems that evolve into contact during the slow mass transfer
phase (case AS). Crosses indicate the systems that could not evolve
into contact while two components are still on MS. We also present
the boundaries (solid lines) for the initial parameters for which
might be the progenitor of VFTS102. The right boundaries are set by
the condition that the system could evolve into contact while two
components are still on the MS. The upper and lower boundaries are
caused by the constraint of the mass of VFTS102. The left boundaries
are not plotted in Fig. 2, which are set by the period ($P_{\rm
ZAMS}$) at which the initially more massive component would just
fill its Roche lobe on the zero-age MS. It is seen from Fig. 2 that
the progenitor binary of VFTS102 with $q_0 \la 0.7$ ($\log q_0 \ga
$0.2) should have an initial orbital period shorter than
$3.76-4.25$\,days and they would evolve into contact during the
rapid mass transfer phase. The progenitor binary of VFTS102 with
$q_0
> 0.7$ ($\log q_0 <$0.2) should have an initial orbital period shorter
than $1.44-1.55$\,days and they would evolve into contact during the
rapid or slow mass transfer phase.

\section{binary population synthesis}
We have performed a series of detailed Monte Carlo simulations to
investigate the binary merger channel. In each simulation, we follow
the evolution of 100 million sample binaries (very wide binaries are
actually single stars) according to grids of stellar models of
metallicity $Z=0.01$ and the evolution channels described above. We
adopt the following input for the simulations: the initial mass
function (IMF) of the primaries, the initial mass-ratio
distribution, and the distribution of initial orbital separations
\citep[e.g.][]{Han 2002, Han 2003, Liu 2009, Wang 2010}:

\begin{itemize}
\item[(i)]
We use a simple approximation to the IMF of \citet{Miller 1979} and
the mass of the primary is generated using a formula of
\citet{Eggleton 1989},
\begin{equation}
M_{10} =\frac{0.19X}{(1-X)^{0.75}+0.032(1-X)^{0.25}},
\end{equation}
where $X$ is a random number uniformly distributed between 0 and 1.
The study of IMF by \citet{Kroupa 1993} supports this IMF.

\item[(ii)]
We take a uniform mass-ratio distribution (Set 1),
\begin{equation}
n(q_{0}) =1, \;\;\;\;  0 \leq q_{0} \leq 1
\end{equation}
where $q_0=M_{20}/M_{10}$ \citep{Mazeh 1992, Goldberg 1994}. In
order to study the influence of the mass ratio distribution, we also
take a rising mass ratio distribution (Set 2)
\begin{equation}
n(q_{0}) =2q_{0}, \;\;\;\;  0 \leq q_{0} \leq 1
\end{equation}
and an alternative mass-ratio distribution where both components are
chosen randomly and independently from the same IMF (Set 3).

\item[(iii)]
We assume that all stars are members of binary systems and that the
distribution of separations is constant in $\log a$ for wide
binaries, where $a$ is the orbital separation and falls of smoothly
at small separation
\begin{equation}
an(a)=\left\{ \begin{array}{ll}
                    \alpha_{\rm sep} (\frac{a}{a_{0}})^m,  & \;\;\;\;\;\;\;\;\;\; \mbox{ $a \leq a_{0}$,} \\
            \alpha_{\rm sep},  & \;\;\;\;\;\; \mbox{ $a_{0} < a < a_{1} $,}
           \end{array}
       \right.
\end{equation}
where $\alpha_{\rm sep}\approx 0.070$, $a_0=10$R$_{\rm \odot}$,
$a_1=5.75 \times 10^6$\,R$_{\rm \odot}= 0.13$\,pc, and $m \approx
1.2$. This distribution implies that the numbers of wide binary
systems per logarithmic interval are equal, and that about 50\% of
stellar systems have orbital periods less than 100\,yr \citep{Han
1995}. A circular orbit is assumed for all binaries.
\end{itemize}

\begin{table}
\begin{footnotesize}
\centering
Table~1.\hspace{4pt} The results of the binary population synthesis for the binary merger channel.\\
\begin{minipage}{85mm}
\centering
\begin{tabular}{c|cccccc}
\hline\hline\
Set&{$n(q_{0})$}&{$N_{\rm m}$}&{$N_{\rm b}$}&{$f_{\rm b}$}&{$N_{\rm s}$}&{$f_{\rm s}$}\\
\hline
1 & ${\rm Uniform}$          &1006 &11581 &8.7\%  &4875  &17.1\% \\
2 & ${\rm Rising}$            &787  &10834 &7.3\%  &4223  &15.7\% \\
3 & ${\rm Uncorrelated}$      &52   &7559  &0.7\%  &4830  &1.1\% \\

\hline
\end{tabular}
\end{minipage}
\end{footnotesize}\\

{Columns: Set-name of different simulation set; $n(q_{0})$-the
distribution of the mass ratio distribution; $N_{\rm m}$-the number
of binaries with total mass $\sim24.5-25.5$\,M$_{\rm \odot}$ that
would merge into single stars; $N_{\rm b}$-the number of the initial
binaries with total mass $\sim24.5-25.5$\,M$_{\rm \odot}$; $f_{\rm
b}$-the fraction of binaries that would merge into single stars
($f_{\rm b}=N_{\rm m}/N_{\rm b}$); $N_{\rm s}$-the number of the
initial single stars with mass $\sim24.5-25.5$\,M$_{\rm \odot}$;
$f_{\rm s}$-the fraction of single stars that might be produced from
the binary merger channel ($f_{\rm s}=N_{\rm m}/(N_{\rm m}+N_{\rm
s})$).}
\end{table}

The results of the binary population synthesis are shown in Table 1
for three sets of simulations. $N_{\rm m}, N_{\rm b}$ and $N_{\rm
s}$ are the number of binaries that would merge into single stars,
the initial binaries and the initial single stars, all of which are
in the range of mass (or total mass for the binaries) from 24.5 to
25.5\,M$_{\rm \odot}$. $f_{\rm b}$ is the fraction of binaries that
would merge into single stars, and $f_{\rm s}$ is the fraction of
single stars that might be produced from the binary merger channel,
where $f_{\rm b}=N_{\rm m}/N_{\rm b}$, and $f_{\rm s}=N_{\rm
m}/(N_{\rm m}+N_{\rm s})$. Set 1 is our standard model for the
binary merger channel and we vary the mass ratio distribution in the
other sets to examine its influence on the final results. The
simulation for the binary merger channel gives $f_{\rm b} \sim
8.7\%$ and $f_{\rm s} \sim 17.1\%$ according to our standard model
(Set 1). This suggests that about 8.7\% of binaries would evolve
into contact and merge into rapidly rotating single stars, and about
17.1\% of single stars might be produced from the binary merger
channel and should have similar properties to VFTS102. The
simulation for a rising mass-ratio distribution (Set 2) gives
$f_{\rm b} \sim 7.3\%$ and $f_{\rm s} \sim 15.7\%$, which are close
to the results in Set 1. If we adopt a mass-ratio distribution with
uncorrelated binary components, two fractions ($f_{\rm b} \sim
0.7\%$ and $f_{\rm s} \sim 1.1\%$) are much lower than the results
in Set 1. This is because most of binary systems have too small a
mass ratio ($q_0$) to locate in the region of the progenitor of
VFTS102 for different initial primaries.

Fig. 3 shows the relative number distribution of the progenitor of
VFTS102 in the initial orbital period$-$mass ratio plane (Set 1). It
is found that the distribution of the progenitor of VFTS102
increases with decreasing orbital period and increasing mass ratio
($q_0)$. This suggests that VFTS102 mainly comes from the binary
with short period and large mass ratio. The most likely progenitor
binary of VFTS102 might have an orbital period $P_{0} < 1.5$\,d and
a mass ratio $q_0 > 0.63$, which roughly corresponds to a binary
with a primary mass ($M_{10} \sim 12.5-15$\,M$_{\rm \odot}$)
according to Fig. 2. This is mainly because the number of initial
binaries increases with the decreasing primary mass according to the
IMF of the primaries, and more systems with low-mass primaries could
be located in the production region of VFTS102.

\begin{figure}
\centerline{\psfig{figure=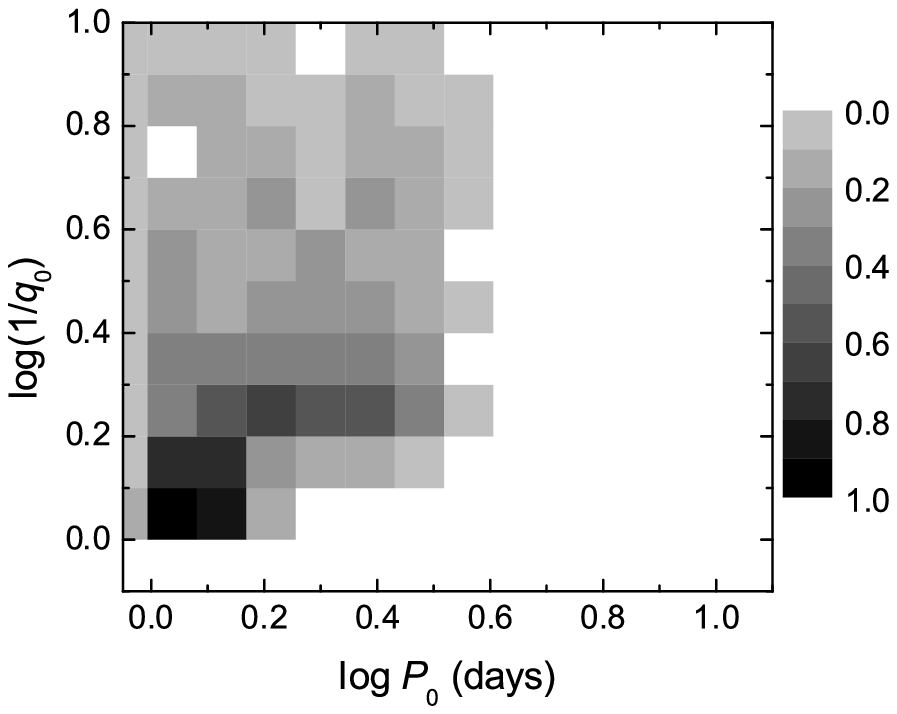,width=7.5cm}} \caption{The
relative number distribution of the progenitor of VFTS102 in the
initial orbital period$-$mass ratio ($\log {P_0}$, $\log (1/q_0)$)
plane.} \label{fig2}
\end{figure}

\section{Discussion and conclusions}
In this paper, we investigated the binary merger channel for the
progenitor of the fastest rotating O-type star VFTS102. We carried
out detailed binary evolution calculations and obtained the initial
parameters of the progenitor of VFTS102 in the orbital period-mass
ratio plane. By performing Monte Carlo simulations, we estimated the
fraction of binaries that would merge into single stars and the
fraction of single stars that might be produced from the binary
merger channel.

\citet{Dufton 2011} proposed that VFTS102 originated from a binary
system, and the mass transfer from its component resulted in the
fast rotation of VFTS102. They suggested that the subsequent
supernova explosion of its component kicked both components and led
to an anomalous radial velocity for VFTS102. We showed that the fast
rotation of VFTS102 could be understood within the framework of the
binary merger channel. In this channel, a detached binary evolves
into contact and then merges into a rapidly rotating single star due
to the orbital angular momentum of the binary systems \citep{Webbink
1976, Jiang 2010, de Mink 2011, Langer 2012}. The anomalous radial
velocity for VFTS102 might be produced by the progenitor binary,
which might be a binary dynamically ejected from a cluster \citep{de
Wit 2005, Eldridge 2011, Gvaramadze 2011}.

We mapped out the initial parameters of the progenitor of VFTS102 in
the binary merger channel. Our results are very similar to those
given by \citet{Pols 1994}. He showed the parameter space of initial
mass ratio and orbital period for the systems with 16\,M$_{\rm
\odot}$, which was explored for the formation of contact binaries.
By performing Monte Carlo simulations, we found that the most likely
progenitor of VFTS102 might have a primary mass $M_{10} \sim
12.5-15$\,M$_{\rm \odot}$, a mass ratio $q_0 > 0.63$ and an orbital
period $P_{0} < 1.5$\,d.

According to the work of \citet{Podsiadlowski 1992}, the fraction of
binaries that experience case A and merge is about 14.8\%, which is
larger than 8.7\% found in this study. This is because they assumed
that case A mass transfer always leads to a merger of two component,
in which some of binaries would reach contact while one or both
components evolve past the terminal MS \citep{Nelson 2001}, and
these binaries are not considered in our study. In addition,
\citet{Langer 2012} suggested that up to 10\% or more of the stars
would be merger remnants, which is close to the result of our study
(17.1\%). Therefore, the binary merger channel might be one of the
main channels for the production of rapidly rotating star like
VFTS102 and it is necessary to consider this channel for the study
of the formation of rapidly rotating stars.

\citet{Tylenda 2011} showed the direct observation evidence that
contact binaries indeed merge into a single object from the
identification of V1309 Sco. The observations also show some
possible progenitor of VFTS102-like objects that are short-period,
massive semi-detached or contact binaries, such as V Pup
\citep[$P$=1.45\,d][]{Andersen 1983}, IU Aur
\citep[$P$=1.811\,d][]{Harries 1998}, V606 Cen
\citep[$P$=1.495\,d][]{Lorenz 1999} and TU Mus
\citep[$P$=1.387\,d][]{Terrell 2003, Qian 2007}. In addition, some
mass should be lost from binary systems during the merger process to
carry away the excess angular momentum \citep{Jiang 2010, Langer
2012}, which might form circumstellar material or a circumstellar
disc. This could explain the observation that VFTS102 has
circumstellar material found by \citet{Dufton 2011}.

In this paper, we investigated the binary merger channel for the
progenitor of VFTS102 without considering the mass loss during the
evolution and merger phase. However, the mass loss is an important
parameter in the evolution, the mass transfer phase and the merger
phase of massive binaries, which needs to be considered in the
future study. In addition, the merger process of the binaries and
the evolution of the merger stars are still uncertain and need to be
investigated further.

\section*{ACKNOWLEDGEMENTS}
It is a pleasure to thank the referee, Professor Ian Howarth, for
his valuable suggestions and comments, which improve the paper
greatly. This work was supported by the Chinese Natural Science
Foundation (10773026, 10821061, 11073049, 2007CB815406, 11033008 and
11103073), the Foundation of Chinese Academy of Sciences
(KJCX2-YW-T24) and the Western Light Youth Project.

\bsp

\label{lastpage}

\end{document}